\documentclass[prl,showpacs,amsmath,amssymb,twocolumn]{revtex4}
\usepackage{mathrsfs}
\usepackage{amsfonts}

\usepackage{amsthm}
\usepackage{dcolumn}
\usepackage{bm}
\usepackage{graphicx}

\newtheorem{lemma}{Lemma}
\newtheorem{theorem}{Theorem}
\newtheorem{corollary}{Corollary}

\newcommand\ket[1]{\ensuremath{|#1\rangle}}

\begin{document}

\title{The Necessary and Sufficient Conditions of Separability for Bipartite Pure States in Infinite Dimensional Hilbert Spaces}

\author{Su Hu}
  \email{hus04@mails.tsinghua.edu.cn}
\author{Zongwen Yu}
  \email{yzw04@mails.tsinghua.edu.cn}
\affiliation{ Department of Mathematical Sciences, Tsinghua
University, Beijing, 100084 China }


\begin{abstract}
In this letter, we present the necessary and sufficient conditions
of separability for bipartite pure states in infinite dimensional
Hilbert spaces. Let $M$ be the matrix of the amplitudes of
$\ket\psi$, we prove $M$ is a compact operator. We also prove
$\ket\psi$ is separable if and only if $M$ is a bounded linear
operator with rank 1, that is the image of $M$ is a one dimensional
Hilbert space. So we have related the separability for bipartite
pure states in infinite dimensional Hilbert spaces to an important
class of bounded linear operators in Functional analysis which has
many interesting properties.
\end{abstract}


\maketitle

A pure state is separable if and only if it can be written as a
tensor product of states of different subsystems. It is also known
that a state $\ket\psi$ of a bipartite system is separable if and
only if it has Schmidt number 1~\cite{NC2000}. Peres presented a
necessary and sufficient condition for the occurrence of Schmidt
decomposition for a tripartite pure state~\cite{Peres1995} and
showed that the positivity of the partial transpose of a density
matrix is a necessary condition for separability~\cite{Peres1996}.
Thapliyal showed that a multipartite pure state is Schmidt
decomposable if and only if the density matrices obtained by tracing
out any party are separable~\cite{Thap1999}. In~\cite{DafaLi2006},
Dafa Li obtained a necessary and sufficient conditions of
separability for multipartite pure state in finite dimensional
Hilbert spaces. In~\cite{YuHu2007}, Yu and Hu gave other necessary
and sufficient conditions, and simplified the proof of the main
result in~\cite{DafaLi2006}. They also obtained an algorithm to
determine the separability of any multipartite pure state
efficiently and quickly. But all of the previous works only aimed to
determine the multipartite pure state in finite dimensional Hilbert
spaces. In this letter, we gave the necessary and sufficient
conditions of separability for bipartite pure state in infinite
dimensional Hilbert spaces and relate this problem to an important
class of bounded linear operators in Functional analysis.

Let $H_{A},H_{B}$ be two infinite dimensional Hilbert spaces. We
define their tensor product as in~\cite{Douglas179}. Let
$H_{A}\stackrel{\displaystyle\otimes}{\scriptscriptstyle a} H_{B}$
denote the algebraic tensor product of $H_A$ and $H_B$ consider as a
linear space over $\mathbb{C}$. It is easy to see $(
\sum_{i=1}^{m}{h_{i}\otimes k_{i}},\sum_{j=1}^{n}{h_{j}^{'}\otimes
k_{j}^{'}})=\sum_{i=1}^{m}{\sum_{j=1}^{n}
{(h_i,h_{j}^{'})(k_i,k_{j}^{'})}}$ defines an inner product of
$H_{A}\stackrel{\scriptstyle{\bigotimes}}{\scriptstyle{a}}H_B$. We
should note that this space is not complete with this inner product.
Pass to completion, we get a Hilbert space. As in Functional
analysis, we call it the tensor product of $H_A$ and $H_B$ and
denote it by $H_A\otimes H_B$. We can prove~\cite{Douglas179}, if
$\left\{e_{\alpha}\right\}_{\alpha\in A}$ and
$\left\{f_{\beta}\right\}_{\beta\in B}$ are orthonormal basis for
$H_A$ and $H_B$ respectively, then $\left\{e_{\alpha}\otimes
f_{\beta}\right\}_{(\alpha,\beta)}\in H_A\otimes H_B$ is an
orthonormal basis for $H_A\otimes H_B$.

Let us now consider two physical systems $A$ and $B$ represented by
the Hilbert spaces $H_A$ and $H_B$ respectively. The joint system is
represented by the Hilbert space $H_A\otimes H_B$. Let $\ket\psi\in
H_A\otimes H_B$ be a pure state of a composite system $AB$. We also
say $\ket\psi$ is separable if and only if it can be written as a
tensor product of states of different subsystems. In this letter, we
suppose $H_A$ and $H_B$ have countable number of dimensions, this is
equivalent to say that $H_A$ and $H_B$ are separable topological
spaces~\cite{Conway}. Let $\ket i$($\ket j$) be the orthonormal
basis for Hilbert space $H_A$($H_B$). From above, we know $\ket
i\ket j$ is an orthonormal basis for Hilbert space $H_A\otimes H_B$.
Since $\ket\psi\in H_A\otimes H_B$, we can give the Fourier
expansion of $\ket\psi$ under this basis. Then we can write
$\ket\psi=\sum_{i,j}{a_{ij}\ket i\ket j}$, where
$a_{ij}\in\mathbb{C}$ and
$\sum_{i=1}^{\infty}{\sum_{j=1}^{\infty}{|a_{ij}|^{2}}}=1$. Let
$M=(a_{ij})$ be the infinite (but countable) dimensional matrix of
the amplitudes of $\ket\psi$. We will prove $M$ is a compact linear
operator in Functional analysis and give the criterion for the
separability.\vspace{3mm}

\noindent
  {\bf{Definition 1.}}~\cite{Conway}
{\em $l^2$ denote the set of all infinite sequences
$\{x_{n}\}_{n=1}^{\infty}$ such that
$\sum_{i=1}^{\infty}{|x_i|^{2}}<\infty$. For $x,y$ in $l^2$ define
the inner product by $(x,y)=\sum_{i=1}^{\infty}{x_{i}\bar{y_j}}$. It
is easy to show $l^2$ is a Hilbert space with this inner product.\/}

\noindent For $x=\{x_n\}_{n=1}^{\infty}\in l^2$, denote
$x=(x_1,x_2,\cdots,x_n,\cdots)^{T},\bar{x}=(\bar{x_1},\bar{x_2},\cdots,\bar{x_n},\cdots)^{T}$.

\begin{theorem}\label{thm:simple}
  \ket{\psi} is separable if and only if there exist two unit vectors $x,y\in l^2$ such
  that $M=xy^{\dagger}$.
\end{theorem}

\begin{proof} $\Rightarrow$ By definition, $\ket{\psi}$ is
separable if and only if we can write $\ket{\psi}=\left(
\sum_{i=1}^{\infty}{x_i\ket{i}}\right)\otimes\left(\sum_{j=1}^{\infty}
{y_j\ket{j}}\right)$ where $\sum_{i=1}^{\infty}{|x_i|^2}=1$ and
$\sum_{j=1}^{\infty}{|y_j|^2}=1$. As above
$\ket\psi=\sum_{i,j}{a_{ij}\ket{i}\ket{j}}$ is the Fourier expansion
of $\ket\psi$ under the orthonormal basis $\ket{i}\ket{j}$ in
$H_A\otimes H_B$. From the definition of tensor product for infinite
dimensional Hilbert spaces. We have
\begin{eqnarray}\label{eq:diag}
   a_{ij}
  &=&
   \left(\ket\psi,\ket{i}\ket{j}\right)\nonumber\\
  &=&
   \left(\left(\sum_{i^{'}=1}^{\infty}
   {x_{i^{'}}\ket{i^{'}}}\right)\otimes \left(
   \sum_{j^{'}=1}^{\infty}{y_{j^{'}}\ket{j^{'}}}\right),\ket{i}\ket{j}\right)\nonumber\\
  &=&
   \left(\sum_{i^{'}=1}^{\infty}{x_{i^{'}}\ket{i^{'}}},\ket{i}\right)
   \left(\sum_{j^{'}=1}^{\infty}{y_{j^{'}}\ket{j^{'}}},\ket{j}\right)\nonumber\\
  &=&
    x_{i}y_{j}
\end{eqnarray}
We set $x=\left(
\begin{array}{c}
  x_1\\
  x_2\\
  \vdots\\
  x_n\\
  \vdots\\
\end{array}\right), y=\left(
\begin{array}{c}
  y_1\\
  y_2\\
  \vdots\\
  y_n\\
  \vdots\\
\end{array}\right)$.
 From $\sum_{i=1}^{\infty}{|x_i|^{2}}=1$ and
$\sum_{j=1}^{\infty}{|y_j|^{2}}=1$, we see $x,y\in l^2$ and as above
$M=\left(a_{ij}\right)= \left(
  \begin{array}{cccc}
    a_{11} &\cdots & a_{1n} & \cdots \\
    \vdots & \ddots & \vdots & \ddots \\
    a_{n1} & \cdots & a_{nn} & \cdots \\
    \vdots & \ddots & \vdots & \ddots
  \end{array}
\right)$. From (1) and the multiplication law for infinite
(countable) dimensional matrices, we see $M=xy^{\dagger}$.

$\Leftarrow$ Suppose that $M=xy^{\dagger},x,y\in l^2$. Because
$M=(a_{ij})$ be the matrix of the amplitudes of $\ket\psi$, we know
$\sum_{i=1}^{\infty}{\sum_{j=1}^{\infty}{|a_{ij}|^{2}}}=1$. From
$M=xy^{\dagger}$ and the multiplication for infinite (countable)
dimensional matrices, we see $a_{ij}=x_{i}\bar{y_j}$. From
$\sum_{i=1}^{\infty}{\sum_{j=1}^{\infty}{|a_{ij}|^{2}}}=1$, we have
$\sum_{j=1}^{\infty}{|y_j|^{2}}\sum_{i=1}^{\infty}{|x_i|^{2}}=1$. We
denote $\parallel x
\parallel=\sum_{i=1}^{\infty}{|x_i|^{2}}<+\infty, \parallel y
\parallel=\sum_{j=1}^{\infty}{|y_j|^{2}}<+\infty$ (because $x,y\in l^2$). Under the
suppose $\widetilde{x}=\frac{x}{\parallel
x\parallel},\widetilde{y}=\frac{y}{\parallel y\parallel }$, we also
have $M=\widetilde{x}\widetilde{y}$. So we can suppose $\parallel x
\parallel =\parallel y
\parallel =1$ and construct two states
$\ket{v}=\sum_{i=1}^{\infty}{x_i\ket{i}},\ket{w}=\sum_{j=1}^{\infty}{\bar{y_{j}}\ket{j}}$.
We have
\begin{eqnarray*}
  \ket{v}\otimes\ket{w}
 &=&
  \left(\sum_{i=1}^{\infty}{x_i\ket{i}}\right)\otimes
  \left(\sum_{j=1}^{\infty}{y_j\ket{j}}\right)\\
 &=&
  \sum_{i=1}^{\infty}{\sum_{j=1}^{\infty}{x_{i}\bar{y_{j}}\ket{i}\ket{j}}}=
  \sum_{i,j}{a_{ij}\ket{i}\ket{j}}=\ket\psi
\end{eqnarray*}
We see $\ket\psi$ is a separable pure state.
\end{proof}

\begin{lemma}\label{lem:simple}
If $M$ is the matrix of the amplitudes of a pure state $\ket\psi$,
then $M=xy^{\dagger}$ if and only if the determinants of all the
$2\times 2$ submatrices of $M$ are zero.
\end{lemma}

\begin{proof}
$\Rightarrow$ $M=xy^{\dagger}$, where $M=(a_{ij})$,
$x=(x_i),y=(y_j)$. As above we see $a_{ij}=x_{i}\widetilde{y_{j}}$.
$m=\left(
\begin{array}{cc}
 a_{il} & a_{ik}\\
 a_{jl} & a_{jk}
\end{array}\right)$ is any $2\times 2$ submatrix of $M$. It is easy to check
$det(m)=a_{il}a_{jk}-a_{ik}a_{jl}=x_{i}\bar{y_l}x_j\bar{y_k}-x_i\bar{y_k}x_j\bar{y_l}=0$.
Therefore if $\ket\psi$ is separable, the determinates of all the
$2\times 2$ submatrices are zero.

$\Leftarrow$ Suppose $M=(M_1,M_2,\cdots,M_j,\cdots)$ (we can suppose
$M_{1}\neq 0$). If $M_1,M_j$ are linearly independent for some
$j>1$, then the $2\times 2$ submatrix is reversible, so $det(m)\neq
0$. This is a contradiction. So for each $j$ we have a constant
$\lambda_{j}\in \mathbb{C}$, such that $M_j=\lambda_{j}M_1$. Then
\begin{eqnarray}\label{eq:diag}
  M&=& (M_1,M_2,\cdots,M_j,\cdots)\nonumber\\
  &=&
  (M_1,\lambda_{2}M_1,\cdots,\lambda_{j}M_1,\cdots)\nonumber\\
  &=&
   M_1(1,\lambda_2,\cdots,\lambda_j,\cdots)
\end{eqnarray}
Set $x=M_1,y=(1,\bar{\lambda_2},\cdots,\bar{\lambda_j},\cdots)^{T}$.
Since $M$ is the matrix of the amplitudes of $\ket\psi$, it is
obviously $x,y\in l^2$. From (2) we have $M=xy^{\dagger}$, as
desired.
\end{proof}

\begin{theorem}\label{thm:simple}
 $\ket\psi$ is a separable pure state if and only if the determinate of all the
$2\times 2$ submatrices of $M$ are zero.
\end{theorem}

\begin{proof}
 This is immediately from theorem 1 and lemma 1.
\end{proof}

\noindent {\bf{Remark}} {\em{Theorem 2 generalize the corresponding
result in~\cite{DafaLi2006} to infinite dimensional Hilbert
spaces.}\/}

\vspace{3mm} \noindent {\bf{Definition 2.}}~\cite{Conway} {\em {An
operator $T$ on Hilbert space $H$ has finite rank if $Im(T)$(the
image of $T$) is finite dimensional.}\/}

\begin{theorem}\label{thm:simple}
$\ket\psi$ is a pure state in $H_A\otimes H_B$. $M$ is the matrix of
the amplitudes of $\ket\psi$, then $M$ is a compact linear operator
on the Hilbert space $l^2$.
\end{theorem}

\begin{proof}
  1) Denote $M=\left(\begin{array}{c}
  M_{1}^{T}\\
  M_{2}^{T}\\
  \vdots\\
  M_{j}^{T}\\
  \vdots
\end{array}\right)$. $\forall x\in l^2$ we have
$$Mx=\left( \begin{array}{c}
 M_{1}^{T}\\
 M_{2}^{T}\\
 \vdots\\
 M_{j}^{T}\\
 \vdots
\end{array}\right)x=\left(
\begin{array}{c}
 M_{1}^{T}\cdot x\\
  M_{2}^{T}\cdot x\\
 \vdots\\
 M_{j}^{T}\cdot x\\
 \vdots
\end{array}\right)=\left(
\begin{array}{c}
 (M_{1},\bar{x})\\
 (M_{2},\bar{x})\\
 \vdots\\
 (M_{j},\bar{x})\\
 \vdots
\end{array}\right)$$

 So $\parallel Mx\parallel^2=\sum_{i=1}^{\infty}{|(M_i,\bar{x})|^2}\leq
 \sum_{i=1}^{\infty}{\parallel M_i\parallel^{2}\parallel x\parallel^{2}}=
 \sum_{i=1}^{\infty}{\sum_{j=1}^{\infty}{|a_{ij}|^2\parallel x\parallel^{2}}}=\parallel x\parallel^2$.
 We get $M$ is a well defined bounded linear operator on $l^2$ with
the norm ${\parallel}M\parallel \leq 1$.

2) To prove $M$ is a compact operator, according to~\cite{Conway} we
should only to show there is a sequence $\{T_n\}$ of operators of
finite rank such that $\parallel M-T_n\parallel\rightarrow
0(n\rightarrow \infty)$. Because $M=(a_{ij})$ is the matrix of the
amplitudes of $\ket\psi$, we have
$\sum_{i=1}^{\infty}{\sum_{j=1}^{\infty}{|a_{ij}|^2}}=1$. We set
$$M_n=\left(
\begin{array}{ccccccc}
  a_{11} & a_{12} & \cdots & a_{1n} & 0 & 0 & \cdots\\
  a_{21} & a_{22} & \cdots & a_{2n} & 0 & 0 & \cdots\\
  \vdots & \vdots & \ddots & \vdots & \vdots & \vdots & \ddots\\
  a_{n1} & a_{n2} & \cdots & a_{nn} & 0 & 0 & \cdots\\
  0 & 0 & \cdots & 0 & 0 & 0 & \cdots\\
  0 & 0 & \cdots & 0 & 0 & 0 & \cdots\\
  \vdots&\vdots&\vdots&\vdots&\vdots&\vdots&\ddots
\end{array}\right),$$
obviously $M_n$ has finite rank.

Denote $s=\sum_{i=1}^{\infty}{\sum_{j=1}^{\infty}{|a_{ij}|^2}}=1,
s_n=\sum_{i=1}^{n}{\sum_{j=1}^{n}{|a_{ij}|^2}}$, from the absolute
convergence of $s$, we know
\begin{equation}
  |s-s_n|\rightarrow 0\quad (n\rightarrow \infty).
\end{equation}

We have
\begin{widetext}
 \begin{equation}
  M-M_n=\left(
  \begin{array}{ccccccc}
   0&0&\cdots&0&a_{1(n+1)}&a_{1(n+2)}&\cdots\\
   0&0&\cdots&0&a_{2(n+1)}&a_{2(n+2)}&\cdots\\
   \vdots&\vdots&\ddots&\vdots&\vdots&\vdots&\ddots\\
   0&0&\cdots&0&a_{n(n+1)}&a_{n(n+2)}&\cdots\\
   a_{(n+1)1}&a_{(n+1)2}&\cdots&a_{(n+1)n}&a_{(n+1)(n+1)}&a_{(n+1)(n+2)}&\cdots\\
   a_{(n+2)1}&a_{(n+2)2}&\cdots&a_{(n+2)n}&a_{(n+2)(n+1)}&a_{(n+2)(n+2)}&\cdots\\
   \vdots&\vdots&\vdots&\vdots&\vdots&\vdots&\ddots
  \end{array}\right)=\left(
  \begin{array}{c}
   B_{1}^{T}\\
   B_{2}^{T}\\
   \vdots\\
   B_{n}^{T}\\
   B_{n+1}^{T}\\
   B_{n+2}^{T}\\
   \vdots
  \end{array}\right).
 \end{equation}
\end{widetext}

From $\sum_{i,j=1}^{\infty}{|a_{ij}|^2}=1$, we see $B_i\in
l^2(i\geq1)$.

 $(l^2)_{1}$ denote the closed unit ball in $l^2$. $\forall \phi\in(l^2)_{1}$, We have
\begin{eqnarray}\label{eq:diag}
 & &(M-M_n)\phi=
 \left(
 \begin{array}{c}
   B_{1}^{T}\\
   B_{2}^{T}\\
   \vdots\\
   B_{j}^{T}\\
   \vdots
  \end{array}\right)\phi\nonumber\\
  &=&\left(
  \begin{array}{c}
   B_{1}^{T}\cdot \phi\\
   B_{2}^{T}\cdot \phi\\
   \vdots\\
   B_{j}^{T}\cdot \phi\\
   \vdots
  \end{array}\right)=\left(
  \begin{array}{c}
    (B_{1},\bar{\phi})\\
    (B_{2},\bar{\phi})\\
    \vdots\\
    (B_{j},\bar{\phi})\\
    \vdots
  \end{array}\right).
\end{eqnarray}

 Then $\parallel (M-M_n)\phi\parallel^2=\sum_{i=1}^{\infty}{|(B_i,\bar{\phi})|^2}\leq
 \sum_{i=1}^{\infty}{\parallel B_i\parallel^2\parallel \phi\parallel^2}\leq \sum_{i=1}^{\infty}
 {\parallel B_i\parallel^2}$. From (4) we get
\begin{eqnarray}\label{eq:diag}
 \sum_{i=1}^{\infty}{\parallel B_i\parallel^2} &=&
  \sum_{i=1}^{n}{\parallel
  B_i\parallel^2}+\sum_{i=n+1}^{\infty}{\parallel B_i\parallel^2}\nonumber\\
  &=&
   \sum_{i=1}^{n}{\sum_{j=n+1}^{\infty}{|a_{ij}|^2}}+\sum_{i=n+1}^{\infty}{\sum_{j=1}^{\infty}{|a_{ij}|^2}}
   \nonumber \\
  &=&|s-s_n|
\end{eqnarray}
\noindent From (5) and (6),
$\parallel(M-M_{n})\phi\parallel^2\leq|s-s_n|,\forall
\phi\in(l^2)_{1}$. From the definition of the norm of the operators
on $l^2$, we have $\parallel
M-M_n\parallel^2=\sum\limits_{\phi\in(l^2)_{1}}{\parallel
(M-M_n)\phi\parallel^2}\leq|s-s_n|$. From (3), we get $\parallel
M-M_n\parallel\rightarrow 0(n\rightarrow \infty)$. So we see $M$ is
a compact operator on $l^2$.
\end{proof}

{\bf{Remark:}} {\em{We know all the compact operators form a closed
two sided ideal in operator algebra and they have many interesting
properties. For example~\cite{Conway}

(1) ``If $A$ is a compact linear operator on $H,\lambda\in
\mathbb{C}$ and $\lambda\neq 0$, then the image of $A$ is closed and
$\dim\ker(A-\lambda I)=\dim\ker(A-\lambda I)^{\dagger}<\infty$",
this is a famous theorem named ``The Fredholm Alternative" in
Functional analysis. Someone call this is ``the linear algebra of
infinite dimensional spaces".

(2) ``$A$ is compact if and only if $A^{\dagger}$ is compact." This
is a theorem of Schaduer.}\/}

\begin{lemma}\label{lem:simple}
  ~\cite{Conway} If $T$ is a positive compact operator,
  then there is a unique positive compact operator $A$ such that $A^2=T$.
   $A$ is called the positive square root of $T$.
\end{lemma}
\begin{lemma}\label{lem:simple}
~\cite{Conway} (Polar decomposition of compact operators.) Let $T$
be a compact operator on Hilbert space $H$ and let $A$ be the unique
positive square root of $T^{\dagger}T$. Then (a) $\parallel
Ah\parallel=\parallel Th\parallel$ for all $h$ in $H$. (b) There is
a unique operator $U$ such that $\parallel Uh\parallel =\parallel
h\parallel$ when $h\perp\ker{T},Uh=0$, when $h\in\ker{T}$ and
$UA=T$.
\end{lemma}
\begin{theorem}\label{thm:simple}
$\ket\psi$ is a pure state in $H_A\otimes H_B$. $M$ is the matrix of
the amplitudes of $\ket\psi$, then $M$ has polar decomposition.
\end{theorem}
\begin{proof}
 Form theorem 3 and Lemma 3.
\end{proof}

We will see if $\ket\psi$ is separable, $M$ is not only a compact
linear operator, but also an operator with rank 1. We know from
Analysis the operators which have finite rank must be a compact
operator~\cite{Conway}.
\begin{lemma}\label{lem:simple}
If $M$ is the matrix of the amplitudes of a pure state $\ket\psi$,
then $M=xy^{\dagger}$, where $x,y\in l^2$ if and only if $M$ is a
bounded linear operator with rank 1.
\end{lemma}
\begin{proof}
$\Rightarrow$ $\forall z\in l^2$, denote $y=\left( \begin{array}{c}
y_1\\
y_2\\
\vdots\\
y_n\\
\vdots \end{array}\right),z=\left(
\begin{array}{c}
  z_1\\
  z_2\\
  \vdots\\
  z_n\\
  \vdots
\end{array}\right)$.

\noindent We have
$$Mz=xy^{\dagger}z=x(\bar{y_1},\bar{y_2},\cdots,\bar{y_n},\cdots)\left(
\begin{array}{c}
  z_1\\
  z_2\\
  \vdots\\
  z_n\\
  \vdots
\end{array}\right)=x\left(\sum_{i=1}^{\infty}{\bar{y_i}z_i}\right),$$
so $Im{M}\subset \textrm{span}{(x)}$. But $M\neq 0$, we get $\dim
Im{M}=1$.

$\Leftarrow$ Suppose $\dim Im{M}=1$. Denote
$M=(M_1,M_2,\cdots,M_i,\cdots)$( we can suppose $M_1\neq 0$).
Suppose the vector $e_i\in l^2$ is the vector with all $0$s except
for a 1 in the $i$th coordinate. We have $Me_1=M_1$ and $Me_i=M_i$.
Then $M_1,M_i\in Im{M}$, but $\dim Im{M}=1$, so there exists
$\lambda_i\in\mathbb{C}$ such that $M_i=\lambda_{i}M_1$. We have
\begin{eqnarray}\label{eq:diag}
 M&=&(M_1,M_2,\cdots,M_i,\cdots)\nonumber\\
 &=&(M_1,\lambda_{2}M_1,\cdots,\lambda_{i}M_1,\cdots)\nonumber\\
 &=&M_{1}(1,\lambda_2,\cdots,\lambda_i,\cdots)
\end{eqnarray}
 Denote $x=M_1,y=(1,\lambda_2,\cdots,\lambda_i,\cdots)^{\dagger}$. $M$ is the
matrix of the amplitudes of $\ket\psi$, we see that $x,y\in l^2$.
Finally, from (7) we get $M=xy^{\dagger}$, as desired.
\end{proof}
\begin{theorem}\label{thm:simple}
  $\ket\psi$ is a separable pure state if and only if $M$ is a bounded linear operator with rank 1.
\end{theorem}
\begin{proof}
  From theorem 1 and lemma 4.
\end{proof}
\begin{corollary}\label{cor:simple}
 $\ket\psi$ is a separable pure state in $H_A\otimes H_B$.
 $M$ is the matrix of the amplitudes of $\ket\psi$. We have following results
  1) $Im{M^{\dagger}}$ is also closed;
  2) $Im{M}=\ker{M^{\dagger}}^{\perp}$;
  3) $Im{M^{\dagger}}=\ker{M^{\perp}}$;
  4) $M$ is an open mapping;
  5) $M^{\dagger}$ is an open mapping.
\end{corollary}
\begin{proof}
 According to theorem 5, $\dim Im{M}=1$, so $Im{M}$ is a closed subspace in $l^2$.
  All the above results follow from the closed range theorem~\cite{Conway}.
\end{proof}

Denote $\mathscr{L}(l^2)$ to be the algebra of bounded linear
operators on $l^2$ and $\mathscr{B}^{0}(l^2)$ to be the ideal of
compact operators on $l^2$. We also define two subsets of
$\mathscr{L}(l^2)$, $\mathscr{F}=\{M\in\mathscr{L}(l^2): M\quad
\textrm{is a bounded linear operator with rank 1}\}$.
$\mathscr{T}=\{M\in\mathscr{L}(l^2): M\quad \textrm{is the matrix of
a}\textrm{ bipatite pure state}\newline \textrm{ of a composite} {
\textrm{system AB}}\}$.

From theorem 3 and 5, we get
$\mathscr{T}\subset\mathscr{F}\subset\mathscr{B}^{0}(l^2)\subset\mathscr{L}(l^2)$,
and relate the separability for bipartite pure states to an
important class of bounded linear operators in Functional analysis.

So, given a pure state $\ket\psi$ in $H_A\otimes H_B$, $M$ is the
matrix of the amplitudes of $\ket\psi$, if $M$ is not a rank 1
operator on $l^2$, we can conclude that $\ket\psi$ is not separable.
But the rank 1 operators in an infinite dimensional Hilbert space
are rare, so the separable pure states in $H_A\otimes H_B$ are also
rare.



\end{document}